\documentclass[printer]{aa} 
\usepackage{graphicx}
\usepackage{txfonts}
\begin{document}
  \title{MHD simulations of radiative jets from young stellar objects:}

  \subtitle{H$\alpha$ emission}

  \author{F.~De~Colle\inst{1} \and A.C.~Raga\inst{2}}

  \offprints{F. De Colle, \\ \email{fdecolle@astroscu.unam.mx}}

  \institute{Instituto de Astronom\'\i a, Universidad Nacional Aut\'onoma de 
             M\'exico, Ap.P. 70264, 04510 DF, Mexico\\ \and
             Instituto de Ciencias Nucleares, Universidad Nacional Aut\'onoma 
             de M\'exico, Ap.P. 70543, 04510 DF, Mexico\\}

  \date{Received xxx / Accepted xxx}

  \abstract{We study the H$\alpha$ emission from jets using two-dimensional 
            axisymmetrical simulations. We compare the emission obtained from 
            hydrodynamic (HD) simulations with that obtained from 
            magnetohydrodynamics (MHD) simulations. The magnetic field is 
            supposed to be present in the jet only, and with a toroidal 
            configuration. The simulations have time-dependent ejection 
            velocities and different intensities for the initial magnetic field. 
            The results show an increase in the H$\alpha$ emission along the 
            jet for the magnetized cases with respect to the HD case. 
            The increase in the emission is due to a better collimation of the 
            jet in the MHD case, and to a small increase in the shock velocity. 
            These results could have important implications for the interpretation 
            of the observations of jets from young stellar objects.

  }

  \maketitle


\section{Introduction}

Collimated outflows are observed in a variety of astrophysical objects, with 
typical spatial scales ranging from $\sim 1$~pc for jets from young stellar 
objects (YSOs) up to several megaparsecs for extragalactic jets. All of these jets 
seem to be associated with accretion disks, which suggests the existence of a 
scale-independent physical mechanism responsible for the ejection and collimation 
of these outflows. The presently more accepted models are the magnetocentrifugal 
models (Blandford \& Payne \cite{BP1982}; Uchida \& Shibata \cite{US1985}), 
in which the ejection is driven by the presence of a dynamically important 
magnetic field in the accretion disk-central object system, and the collimation 
of the jet is due to the toroidal component of the magnetic field, which is able 
to collimate the outflows by pinching forces. The toroidal magnetic field is 
generated by the twisting of the magnetic field due to the rotation of the system.
The region where this process acts is too close to the central object to be 
resolved observationally, and one possible way to obtain some insight into this 
region is by studying the properties of the outflows.

In particular, a lot of progress has recently been made regarding observations of 
the outflows from YSOs (see the review by Reipurth \& Bally
\cite{RB2001}). 
The caracteristic spectral emission of these objects
is believed to come from the region behind the shock, from
recombination of the ionized gas (for the hydrogen lines)
and electron excitation (and de-excitation) within ions (Schwartz
\cite{S1975}), 
and the typical knot structure visible along the jet could be
interpreted as due to a time-dependent ejection from the young stellar 
object (YSO) (Reipurth \cite{R1989}; Raga et al.\cite{Ral1990}).

In the past few years, several authors have studied the effect of the magnetic 
field on the dynamical evolution of HH objects (e.~g. O'Sullivan \& Ray 
\cite{OR2000}; Stone \& Hardee \cite{SH2000}; Cerqueira et al. \cite{Cal1997}).
Frank et al. (\cite{Fal1999}) showed that ambipolar diffusion could be important 
to smear out the magnetic field ejected with the jets, but only on timescales 
comparable or larger than the dynamical timescales of the jets. 
The determination of a magnetic field in the outflows would
represent a test for the magnetocentrifugal mechanism. In particular,
the presence of a dynamically important toroidal component of the
magnetic field would represent an indirect proof of this mechanism.
However, direct observations of magnetic fields in jets are very
difficult, and there is yet no clear observational determination of 
the magnetic field intensity in jets.

Several hydrodynamic (HD) simulations with calculations of the spectral emission of 
HH objects have been presented in the past (e.~g. Blondin et al. \cite{Bal1990}; 
Raga \cite{R1994}), but such calculations have never been published for the 
magnetized case. The simulations with magnetic fields have concentrated on the 
dynamical aspects and the evolution of the jet rather than on obtaining predictions 
of the emitted spectrum. 
These simulations usually include a radiative cooling rate (given, e.g., by the 
coronal cooling function of Dalgarno \& McCray \cite{DM1972}), and different 
magnetic field configurations. The main features found in magnetized jets with 
respect to HD jets are the presence of a ``nose cone'', better collimation, an 
increase in the density along the direction of propagation, and some effects on 
Kelvin-Helmoltz and Rayleigh-Taylor instabilities (resulting in changes in the 
leading bow shock, e.~g. Todo el al \cite{Tal1993}; Cerqueira \& de Gouveia 
Dal Pino \cite{CD1999}).

With respect to the emitted spectrum, Hartigan, Morse \& Raymond (\cite{Hal1994}) 
compared observed and predicted emission lines ratios (using plane-parallel shock 
models) to find an upper limit of 30 $\mu$G for the magnetic field of the jet. 
Cerqueira \& de Gouveia Dal Pino (\cite{CD2001a}), using a semiempirical formula 
(valid for a shock velocity between $20$ and $80$ km s$^{-1}$) found the ratio 
between the H$\alpha$ emission of a magnetized and a non-magnetized jet.
They obtained that the H$\alpha$ emission increases due to the presence of a magnetic
field, and that the dominant cause of this increase is the toroidal component of the 
magnetic field.

Therefore, no direct calculation of the effect of the magnetic field on the emission 
from a jet has yet been made. Trying to fill this gap, we have carried out 2D, 
axisymmetric, MHD simulations of jets, looking for the differences in the predicted 
H$\alpha$ emission from magnetized jets with respect to the hydrodynamic case. 
The paper is organized as follows. In section 2, we explain in some detail the 
numerical algorithm that has been used, the initial conditions of the simulations and 
the approximations used to calculate the emission. In section 3 we summarize and 
discuss the results obtained, and in section 4 we draw our conclusions.


\begin{figure*}
  \centering
  \includegraphics[width=18cm]{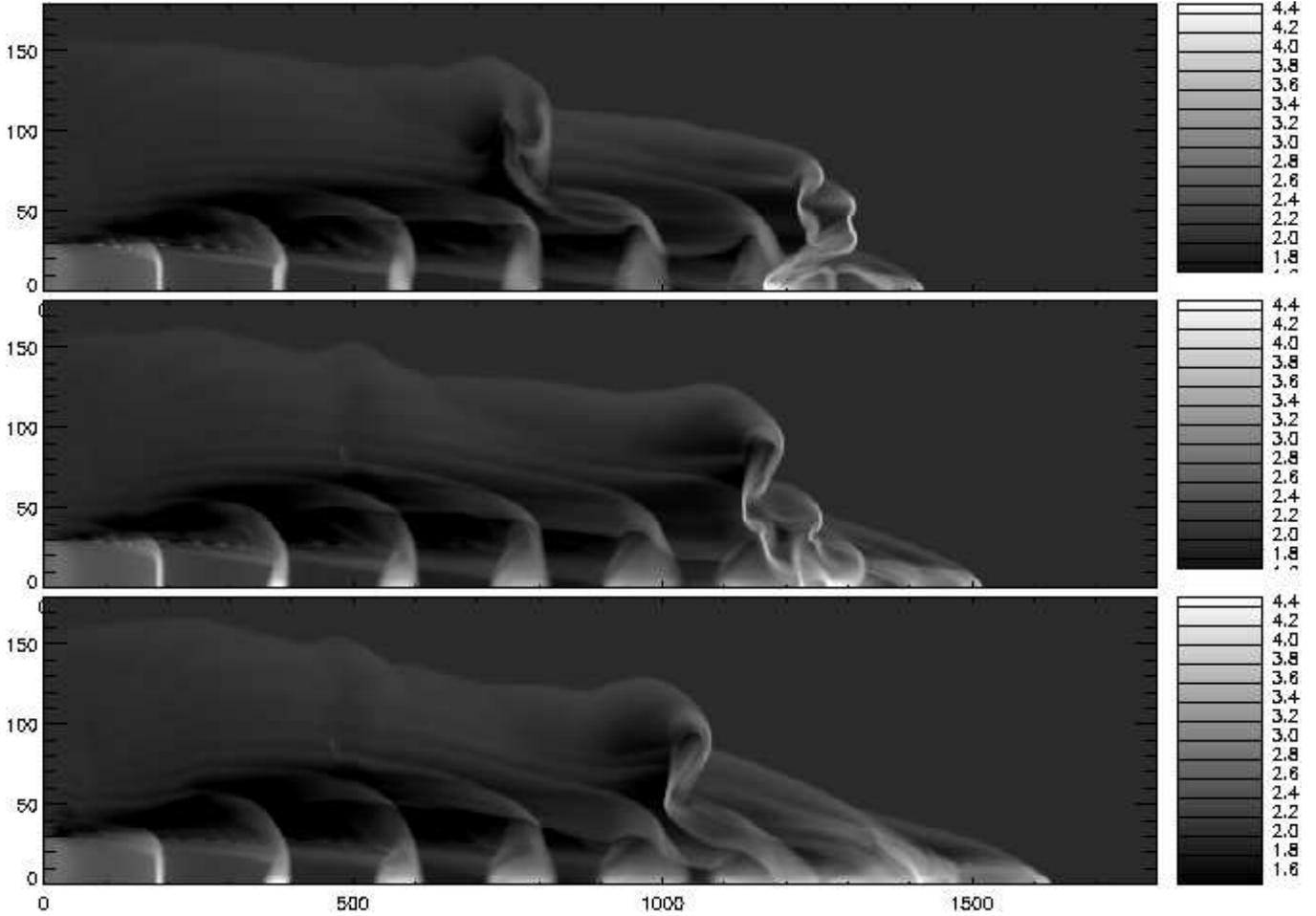}
  \caption{Numerical density for the models HD (hydrodynamic), weak MHD ($\beta=1$), 
           and strong MHD ($\beta=0.4$), after a 500 yrs integration time. 
           The right bar gives the logarithm of the numerical density (in cm$^{-3}$)
           The axes are labeles in pixels, and the displayed domain has a physical
           size of ($L_z$,$L_r$) = ($3\times 10^{17}$, $3\times 10^{16}$) cm.
           A pixel correspond to a physical size of $1.66 \times 10^{14}$ cm.}
  \label{fdens}
\end{figure*}



\section{Simulations}

\subsection{Numerical method}

To study the H$\alpha$ emission from YSO jets, we have carried out a set of 2D,
axysimmetrical simulations using a MHD code, modified to include the cooling and 
the evolution of the hydrogen ionization fraction. 
We solve the following set of equations:

\begin{equation}
  {\partial \rho \over\partial t}
    +\nabla\cdot(\rho \vec v)  = 0 
  \label{mhd1}
\end {equation}
\begin{equation}
  {\partial \rho \vec v \over\partial t}
    +\nabla\cdot
    (\rho \vec v \vec v + p_{tot} I-\vec B \vec B)=0
  \label{mhd2}
\end {equation}
\begin{equation}
  {\partial e \over\partial t}
    +\nabla\cdot
    ((e+p_{tot})\vec v-(\vec v \cdot \vec B)\vec B) =-n^2 \Lambda
  \label{mhd3}
\end {equation}
\begin{equation}
  {\partial \vec B \over\partial t}
    +\nabla\cdot
    (\vec v \vec B - \vec B \vec v) =0
  \label{mhd4}
\end {equation}
\begin{equation}
  {\partial n_{H^0}\over\partial t}+\nabla \cdot (n_{H^0} \vec v)=
          n_{H^+}n_e\alpha(T)-n_{H^0}n_e C(T)
  \label{mhd5}
\end{equation}
where $\rho$ is the mass density, $\vec v$ is the velocity vector, 
$p_{tot}=p_{gas}+B^2/2$ is the (magnetic $+$ thermal) total pressure, 
$I$ is the identity matrix, $\vec B$ is the magnetic field normalized 
with respect to $\sqrt{4\pi}$, $e$ is the total energy defined as 
$e= \frac{1}{\gamma-1}p_{gas}+ \frac{1}{2} \rho v^2+ \frac{1}{2} B^2$ 
(with $\gamma=5/3$), and finally $\Lambda$ is the cooling function.
These equations represent the conservation of mass (\ref{mhd1}), 
momentum (\ref{mhd2}), energy (\ref{mhd3}) and magnetic flux (\ref{mhd4}).
Eq.~(\ref{mhd5}) represents the evolution of the hydrogen neutral fraction.
This equation is coupled to the others by the cooling function $\Lambda$, 
which is a function of the ionization fraction.
In Eq.~(\ref{mhd5}):
\begin{displaymath}
  \alpha(T)=2.55 \cdot 10^{-13} (T/10^4)^{-0.79}
\end{displaymath}
\begin{displaymath}
  C(T)=5.38 \cdot 10^{-11} e^{-\frac{157828}{T}}/\sqrt T
\end{displaymath}
are the recombination and collisional ionization coefficients (Cox
\cite{C1970}), and $T$ is the gas temperature.

To solve this equation system, we use a second order upwind scheme, which integrates 
the MHD equations using a Godunov method with a Riemann solver.
The Riemann problem is solved using primitive variables and the magnetic field 
divergence is maintained close to zero using the CT method (Toth \cite{T2000}).
The algorithm is similar to the one of Falle, Komissarov and Joarder (\cite{Fal1998}), 
except that to preserve a small divergence of the magnetic field we use a constrained 
transport method.
However, the magnetic field divergence is automatically equal to zero in all our
axysimmetrical simulations because we use a toroidal magnetic field configuration.
To treat correctly the source terms in cylindrical coordinates, these terms are
averaged on the cell volume (see the discussion in the appendix of Falle \cite{F1991}).
The code was tested with one- and two-dimensional tests (Ryu et al. \cite{Ral1995}; 
Toth \cite{T2000}; also see De Colle \cite{D2005}\footnote{A detailed description of
the free distributed numerical code with several different one, two
and three dimensional numerical tests will
be soon available at www.astroscu.unam.mx/\~{}fdecolle}).

To solve Eq.~(\ref{mhd3}), we first integrate the equation without the cooling term
finding a new value for the energy $e'$, and afterwards we use this new energy value to
integrate the equation
\begin{equation}
  {\partial e' \over \partial t} = -n^2 \Lambda (T) = -n^2 \Lambda_0 {T \over T_0},
  \label{energy}
\end{equation}
where $n$ is the number density (assumed to be constant within the timestep)
and we approximate $\Lambda(T)$ in a locally linear form.
This equation has an exact solution for the gas pressure:
\begin{equation}
  p_{gas}=p_{0,gas} e^{-2 \Lambda_0 n_H \Delta t/3 k_B T }.
  \label{pgas}
\end{equation}
where $n_H$ is the total (ionized more neutral) hydrogen density, $T$ is the gas
temperature, $k_B$ is the Boltzmann constant and $\Delta t$ is the time step. 
A floor of 1000 K is used as the minimal temperature value.

Tests for adiabatic and radiative magnetized and HD jet propagation (reproducing the 
results of Frank et al. \cite{Fal1998}) are reported in a previous paper 
(De Colle \& Raga \cite{DR2004}).

\subsection{Cooling term}

We have used a non-equilibrium cooling function considering the energy loss due
to collisional excitation of oxygen, radiative recombination of hydrogen, collisional
ionization of the hydrogen and excitation of Lyman-alpha line (Biro et al. \cite{Bal1995}).
If one computes an coronal equilibrium cooling curve using this simplified scheme,
one obtains a cooling function which is similar to the one of 
Dalgarno \& McCray (\cite{DM1972}), which is usually used in MHD jets simulations.
For temperatures between 10$^4$ K and 10$^5$ K it is necessary to use a
non-equilibrium function instead of the simpler equilibrium cooling function.
However, using our non-equilibrium cooling function, we find that the dynamical
evolution of the jet and the main features (such as better collimation, presence 
of a small nose cone, etc.) are very similar to the ones of previous simulations 
(obtained with a coronal equilibrium cooling function).

To obtain the H$\alpha$ emission we consider the contributions from the
radiative recombination cascade (Aller \cite{A1984})
and collisional excitations from the $n=1$ state (Giovanardi and Palla \cite{GP1989}).
The H$\alpha$ emission is then obtained as a function of the local temperature
of the gas $T$, and of the neutral and total densities $n_{H^0}$ and $n_H$.

\subsection{Initial conditions}


\begin{figure*}
  \centering
  \includegraphics[width=18cm]{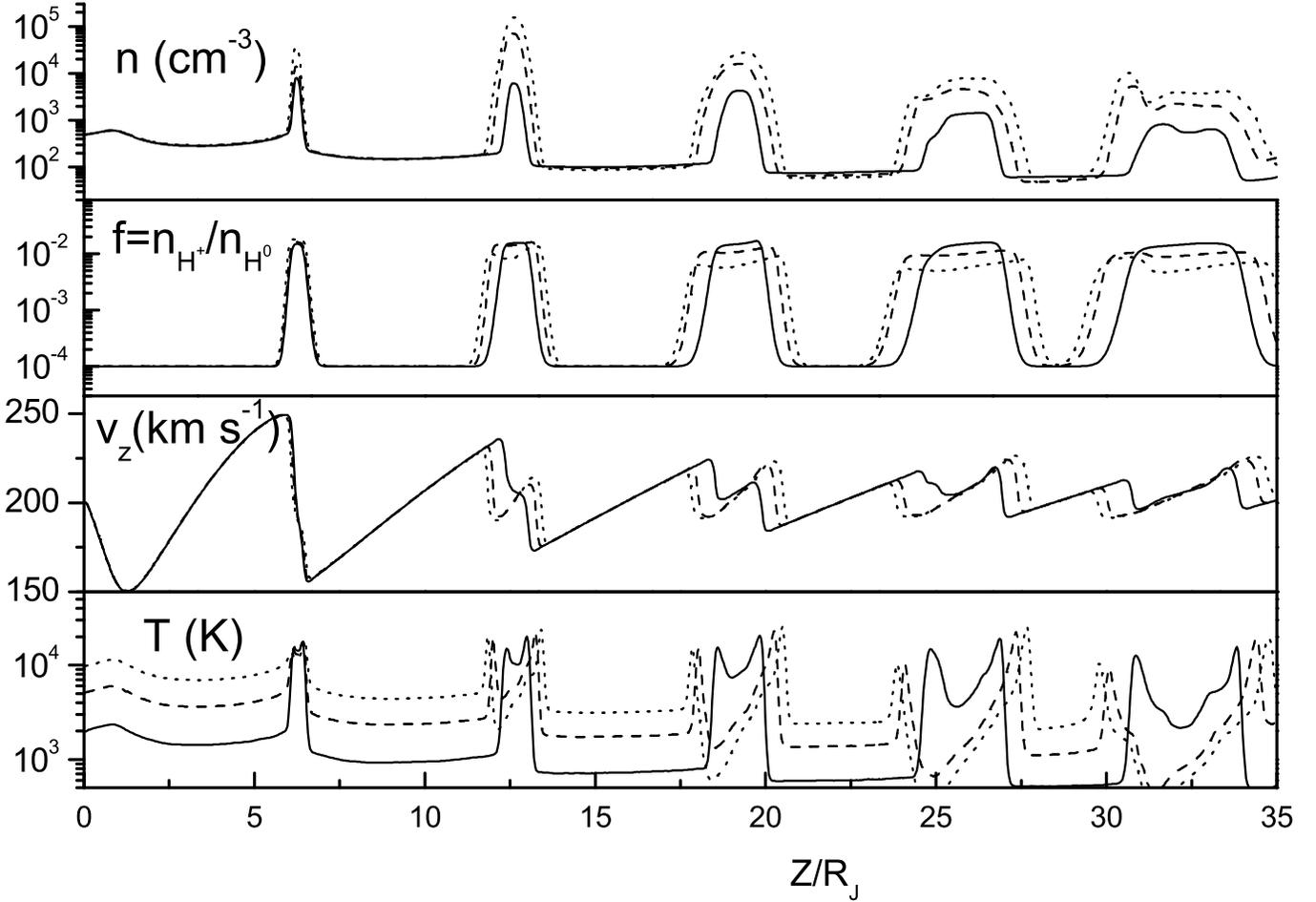}
  \caption{Axial cuts of the density (top), hydrogen ionization fraction,
           axial velocity and temperature (bottom) for the different models. 
           The solid lines correspond to the HD model, the dashed lines 
           correspond to the weak magnetic field ($\beta=1$) model, and the 
           dotted lines correspond to the strong magnetic field ($\beta=0.4$) 
           model.}
  \label{fprof}
\end{figure*}


In our numerical simulations, we use a uniform grid of axial/radial size
($L_z$,$L_r$)=($3\times 10^{17}$, $3\times 10^{16}$) cm.
The resolution is of $1.67\times 10^{14}$ cm, corresponding to $1800 \times 180$ cells
along the $z$- and $r$-axes, respectively.

In the models, the atomic jet has an initial radius of
$6\times 10^{15}$~cm (corresponding to 30 cells) and is moving in the $z$-direction
with a mean velocity $v_0=200$ km s$^{-1}$ (on this mean velocity, we superimpose
a sinusoidal velocity variability, see below).
The initial jet density in all cases is $n_{jet}=500$ cm$^{-3}$, the ambient density is
uniform, with a value $n_{amb}=100$ cm$^{-3}$, and the ambient sound speed
is $c_s=10$ km s$^{-1}$ (corresponding to $T=10000$~K).
The edge of the jet and the ambient medium are initially in radial pressure
equilibrium. The initial jet temperature has a radial profile which is discussed
below.

The jet evolution is followed for a maximum of 500~yr as it propagates along the 
$z$-direction. We use reflection boundary conditions 
at $z$=0 and $r$=0, and open boundary conditions for the outer boundaries of the
$x$ and $z$-axes.

To obtain a number of knots along the direction of propagation of the jet,
we impose a sinusoidally variable ejection velocity of the form:
\begin{equation}
  v(t)=v_0 \left(1+A\sin\frac{2\pi t}{\tau}\right)
  \label{varvel}
\end{equation}
where $A = 0.25$ and $ \tau = 50$ yr are the amplitude and the period of the 
perturbation in the velocity.

Following Lind et al. (\cite{Lal1989}), we use the simple profile for the toroidal 
magnetic field:
\begin{equation}
  B(r)=\left\{ \begin{array}{cc} 
  B_m\frac {r}{R_m} & 0\leq r<R_m \\
  B_m\frac {R_m}{r} & r_m\leq r<R_j \\
  0 & R_{j}\leq r \end{array} \right.
  \label{lind}
\end{equation}
To ensure initial hydromagnetic equilibrium in the radial direction, it is necessary 
that the magnetic and thermal pressures satisfy the radial equilibrium equation:
\begin{equation}
  \frac{dp}{dr}+\frac{B}{r}\frac{d(rB)}{dr} = 0.
  \label{equil}
\end{equation}
Integrating this equation, using the magnetic field profile given by Eq.
(\ref{lind}), it is possible to find the thermal pressure profile (Lind et al. 
\cite{Lal1989}; O'Sullivan \& Ray \cite{OR2000}):

\begin{equation}
  p(r)=\left\{ 
  \begin{array}{cc}
    (\alpha+\frac{2}{\beta}(1-\frac{r^2}{R_m^2})) p_{amb} & 0\leq r<R_m \\ 
     \alpha p_{amb} & r_m\leq r<R_j \\ 
     p_{amb} & R_{j}\leq r
  \end{array}
  \right. 
  \label{pressure}
\end{equation}
where $\alpha$ is the integration constant of Eq.~(\ref{equil}), which is
connected to $R_m$ by the relation:

\begin{equation}
  \alpha=1-\frac{1}{\beta_m}\left(\frac{R_m}{R_j}\right)^2\,.
  \label{alpha}
\end{equation}
This is the magnetic field profile used by Lind et al. (\cite{Lal1989}),
O'Sullivan \& Ray (\cite{OR2000}) and Frank et al. (\cite{Fal1998}), and is similar 
to the one Stone \& Hardee (\cite{SH2000}, who use a slightly different form for 
the magnetic field profile).

The free parameters in these formulae are the values of $R_m$ (or equivalently 
$\alpha$). We choose $R_m$=0.6$R_j$. 
In this way the $\alpha$ parameter (obtained from Eq. \ref{alpha}), is
always positive and the thermal pressure (obtained from Eq. \ref{pressure}) 
is also positive.


\begin{figure*}
\centering
  \includegraphics[width=18cm]{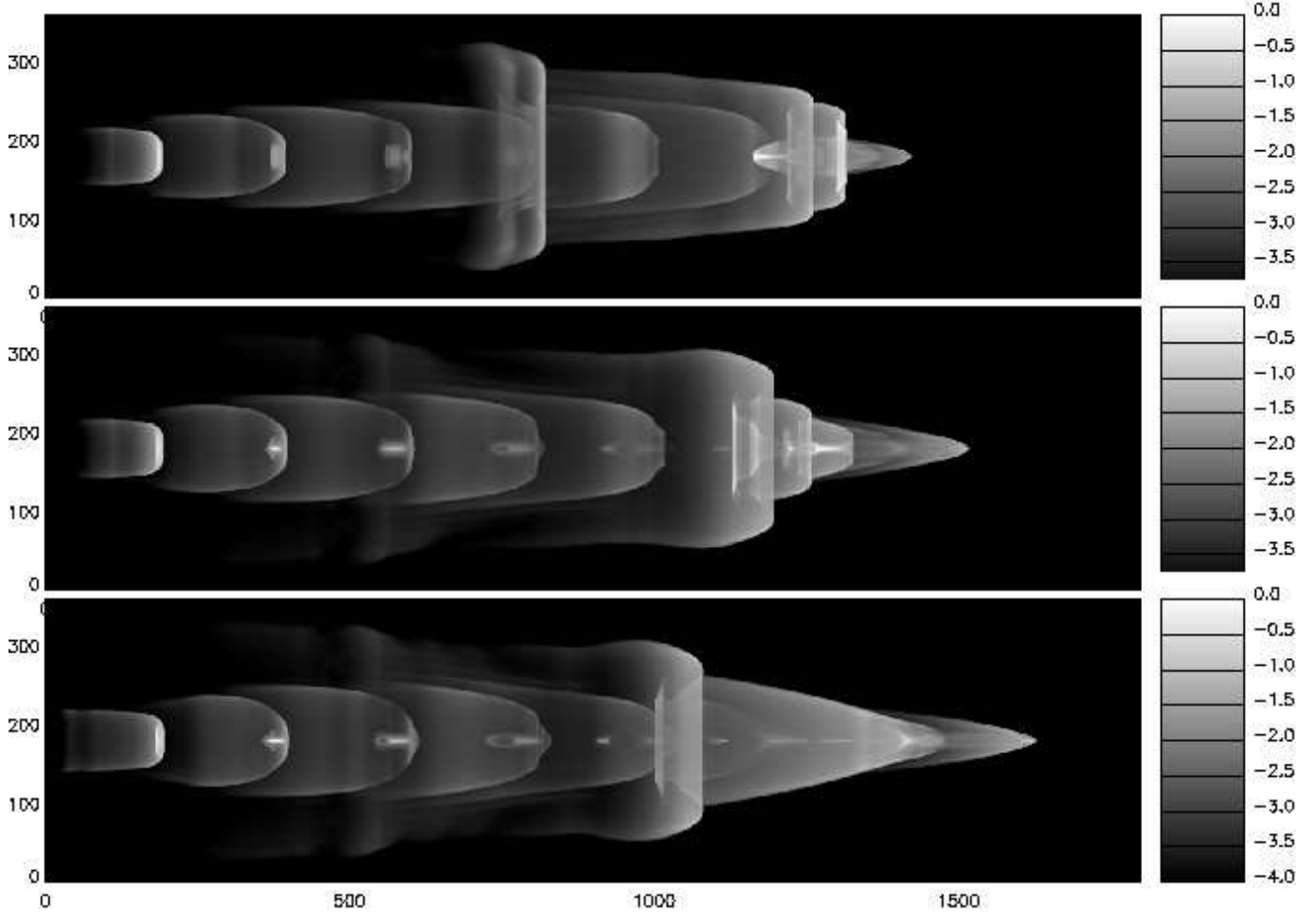}
  \caption{H$\alpha$ emission maps for the three computed models.
           The frames correspond to the HD, weak MHD ($\beta=1$), 
           and strong MHD ($\beta=0.4$) models (up to bottom),
           after 500 yrs integration time. The right bar gives the logarithm of the 
           H$\alpha$ emission, normalized to the maximum value of each frame.
           The axes are labeles in pixels, and the displayed domain has a physical
           size of ($L_z$,$L_r$) = ($3\times 10^{17}$, $3\times 10^{16}$) cm.
           A pixel correspond to a physical size of $1.66
           \times 10^{14}$ cm.}
  \label{fhalf}
\end{figure*}


We choose a toroidal field geometry for two reasons. The first one is that
a toroidal magnetic field is necessary to collimate the jet during the
ejection-collimation process, and would then also have to be present in the 
propagating jet.
The second reason is that in shocks, the ``important'' component of the magnetic field is
the one parallel to the surface of the shock (which is the toroidal component in our geometry).
In fact, all of the previous simulations confirm this idea. Stone \& Hardee (\cite{SH2000})
and Frank et al. (\cite{Fal1998}) obtained relevant differences in MHD jets (with respect 
to the HD case) with a strong toroidal component of the magnetic field, and less 
important differences for a longitudinal magnetic field 
(see also Gardiner et al. \cite{Gal2000}).
Cerqueira \& de Gouveia Dal Pino (\cite{CD2001a}, \cite{CD2001b}) also showed that larger effects
in the calculated H$\alpha$ emission are also obtained for toroidal field configurations.

We have computed models with different magnetic field intensities,
ranging from $B_m=0$ to $B_m=93.2$ $\mu$G (see equation \ref{lind}),
corresponding to $\beta=\infty$ to $\beta=0.4$, (where $\beta=2P_{gas}/B^2$).
In all of our simulations, the ambient medium is not magnetized.
The sound velocity and the Alfv\'en velocities, at the centre
  of the initial cross section, ranges from
  c$_{s,0}=$c$_{s,amb}/\sqrt{5}\approx 4.5$ km/s and c$_A=0$ (for the 
$\beta=\infty$ case) to  c$_{s,0} \approx 10$ km/s and c$_A\approx$
  17.3 km/s (for the $\beta=0.4$ case).


\section{Results}
\subsection{General Results}

The stratification of the numerical density for simulations with different values
of the magnetic field intensity are shown in Fig. 1 for a $t=500$~yr time integration.
The top frame shows the HD simulation, the central frame a weak magnetic field
simulation (with $\beta$=1, $B_m=58.9\,\mu$G) and 
the bottom frame a strong magnetic field simulation ($\beta$=0.4, $B_m=93.2\,\mu$G).

The results show an increasing collimation for increasing toroidal magnetic fields.
The knots are due to the imposed variable injection velocity (see Eq. \ref{varvel}).
The head of the jet presents a very complicated structure, due to the presence
of instabilities (see, e.~g., Blondin et al. 1990) and to the interaction of successive
knots which catch up with the head.

An important result is that the H$\alpha$ emission of the leading head is strongly
dependent on $\beta$. This is a result of the fact that our magnetized jet
simulations develop "nose cones". These structures are of somewhat dubious
reality, as they might disappear in 3D jets without perfect
axisymmetry (see Cerqueira \& de Gouveia Dal Pino \cite{CD2001b}).

The structure of the head is strongly time-dependent,
and will not be considered in the following analysis.
In Fig.~\ref{fprof} we show the density, the ionization fraction, the axial flow velocity and the
temperature as a function of position along the symmetry axis for the three
computed models. This figure does not include the region of the head of the jet.

Some effects due to the presence of the magnetic field are evident. 
We see that the on-axis ($r=0$) density 
increases $\sim$~one order of magnitude from the HD to the $\beta$=0.4 model.
The initial temperature is bigger for the magnetized cases due to the different pressure
profiles present in the initial configuration of the jet (which correspond to
temperature profiles, as the density is assumed to be constant, see section 2) . However,
within the internal working surfaces we obtain lower temperatures as a function
of increasing magnetic field intensity (similar results were obtained by
Stone \& Hardee \cite{SH2000}). The ionization fraction in the central regions of the knots
decreases as a function of increasing magnetic field strength (which is consistent
with the temperature decrease described above).

The separation between the two shocks of the knots increases both as a function
of distance from the source and of increasing magnetic field intensity. This latter
effect is due to the fact that the magnetic field tends to stop the lateral expansion
of the material within the knots, so that more material remains in the region close to the
symmetry axis. With respect to the velocity, we see that the shock velocities
(which correspond to the velocity jumps observed at the edges of the knots
in Fig. 2) increase as a function of increasing magnetic field. 
Finally, the propagation velocity of the head of the jet increases substantially
as a function of increasing magnetic field (as can be seen in Fig. 1). This effect
is also due to the increased collimation obtained for the MHD jets.

\subsection{Intensity maps}
From our models, we have computed H$\alpha$ emission line maps from the models.
These maps are obtained integrating the emission coefficient along lines of sight,
and are computed assuming an optically thin jet moving parallel to the plane of the sky.
In Fig. 3 we show the H$\alpha$ emission line maps predicted from the models
presented in Fig. 1. It is evident that the H$\alpha$ emission is more concentrated
to the jet axis in the magnetic jet models than in the HD jet.

To have a quantitative value of the H$\alpha$ emission from the knots, we integrate
the emission over regions around each knot. To this effect, we integrate the emission
within circular diaphragms with a radius of $7\times 10^{15}$~cm centered on the
successive knots. In fact, the regions outside these diaphragms have negligible
emission (see Fig. 3).

In Fig. 4, we show the results obtained for the different models. The points represent
the ratio of the H$\alpha$ luminosities of the successive knots with respect to the
H$\alpha$ lumninosity of the corresponding knot in the HD
simulation. For the first knot (located at $z\approx 4R_J$) we find
that all of the magnetized jet models have almost the same H$\alpha$
luminosity as the HD jet model (i. e.,
$L_{H\alpha}(MHD)/L_{H\alpha}(HD)\approx 1$). The value
of $L_{H\alpha}(MHD)/L_{H\alpha}(HD)$ increases as a function of
$z$ in the second and
third knots (located at $z\approx 8$ and 14$R_J$) and  
then decreases again for larger $z$ values (i. e., for the
fourth and fifth knots, located at $z\approx 19$ and $25 R_J$).

From this Fig. 4,
we also see that a model with $\beta=5$ ($B=26.3\mu$G) produces knots with H$\alpha$ luminosities
which are basically indistinguishable from the HD simulation. For higher magnetic
fields (i.~e., for lower values of $\beta$), knots with increasingly larger H$\alpha$
luminosities (compared to the HD case) are obtained. The largest increase is found
for the knot at position $z\approx 14 r_j$ ($\approx 8.5\times 10^{16}$~cm)
These results agree at least qualitatively with the ones shown in
Fig. 3 of Cerqueira and de Gouveia Dal Pino (\cite{CD2001a}), who find increases in the
H$\alpha$ luminosity by factors of $\sim 3$-4 with respect to the HD case for
a $\beta=1$ magnetized jet model.


\begin{figure}
   \centering
   \includegraphics[width=8cm]{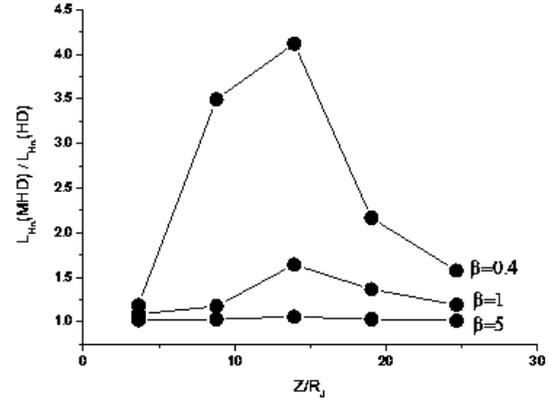}
  \caption{Ratios of the H$\alpha$ luminosities of the knots along magnetized jet models
           with respect to the H$\alpha$ luminosity of the corresponding knots in a HD jet,
           as a function of knot position along the jet axis.
           These luminosities have been obtained by integrating the H$\alpha$ intensity
           over circular diaphragms centered on the successive knots (see the text and Fig. 3).}
\end{figure}



\section{Conclusions}

We have presented 2D numerical simulations of HD and magnetized, variable
jets propagating in a homogeneous medium. Models with increasing magnetic
fields show an increase in the H$\alpha$ luminosities of the successive knots
(which correspond to internal working surfaces which result from the injection
velocity variability). This result confirms the work of Cerqueira \& de Gouveia Dal Pino
(\cite{CD2001a}, \cite{CD2001b}), who estimated the H$\alpha$ luminosity of the clumps along MHD
jet simulations using a fit to predictions of plane-parallel shock models.

Somewhat surprisingly, our work presents the first predictions of emission
line maps from MHD HH jet models. Therefore, our calculations for the first
time show the emission line morphologies that would be expected for such models.

We find that the H$\alpha$ emission of the leading head of the jet differs
quite strongly between the HD and MHD cases. This is a result of the fact
that our simulations develop extended ``nose cones'' (of somewhat dubious
reality, as these structures might disappear in 3D jets without perfect
axisymmetry, see Cerqueira \& de Gouveia Dal Pino \cite{CD2001b}).

For the knots along the jet, we find that for increasing magnetic field strengths
we obtain emission structures with stronger peaks towards the symmetry axis.
This can be clearly seen in Fig. 3, in which the $\beta=0.4$ model has knots
which are dominated by an elongated emission component along the jet axis.

This different type of knot morphology is interesting in terms of observations of
HH jets. It has been a long-standing fact that while some HH jets (notably
HH~111, see e.~g. Reipurth et al. \cite{Ral1997}) show compact knots with 
``bow shock-like'' morphologies which resemble the predictions from HD variable 
jet models (see Masciadri et al. \cite{Mal2002}), other HH jets 
(e.~g., HH~30, see Lopez et al. \cite{Lal1995}) have emission knots with 
axially elongated structures. 
This second kind of morphology could not be modeled successfully in
terms of variable HD jet models, and suggested the presence of a different
mechanism for knot formation. We now find that variable jets with a strong enough
toroidal magnetic field do lead to the formation of axially elongated knots, which
in principle might be used to model objects such as HH~30.

The present paper is limited to a study of the effect of a toroidal magnetic
field on the H$\alpha$ emission of variable jet flows. In a future paper, we
will present a study of a more extended set of emission lines. From such
a study, we will attempt to produce a set of line diagnostics which could
be used to estimate the magnetic field strength along observed HH jets.


\begin{acknowledgements}
This work was supported by the CONACyT grants 43103-F and 41320 and the DGAPA 
(UNAM) grant IN 112602. 
FDC acknowledges support from a fellowship of the DGEP-UNAM.
\end{acknowledgements}


\end{document}